\documentclass[graybox]{svmult}

\usepackage{mathptmx}       
\usepackage{helvet}         
\usepackage{courier}        
\usepackage{type1cm}        
\usepackage{makeidx}         
\usepackage{graphicx}        
\usepackage{multicol}        
\usepackage[bottom]{footmisc}

\usepackage{bm}

\usepackage{hyperref} 

\makeatletter
\providecommand*{\toclevel@author}{0}
\providecommand*{\toclevel@title}{0}
\makeatother

\makeindex             

\newcommand{\params}{\mathcal{P}}
\newcommand{\like}{\mathcal{L}}
\newcommand{\Dobs}{D_{\rm obs}}

\newcommand{\scond}[1]{\qquad\quad||\;#1}

\long\def\symbolfootnote[#1]#2{\begingroup%
\def\thefootnote{\fnsymbol{footnote}}\footnote[#1]{#2}\endgroup}

\begin{document}

\title*{Bayesian astrostatistics:  a backward look to the future$^\dagger$}
\titlerunning{Bayesian astrostatistics}
\author{Thomas J. Loredo}
\institute{Thomas J. Loredo \at Center for Radiophysics \& Space Research,
Cornell University, Ithaca, NY 14853-6801, \email{loredo@astro.cornell.edu}\\
\\
$^\dagger$This paper is a lightly revised version of an invited chapter for
{\em Astrostatistical Challenges for the New Astronomy} (Joseph M. Hilbe, ed.,
Springer, New York, 2012), the inaugural volume for the Springer Series in
Astrostatistics.  The volume commemorates the first invited session
on astrostatistics held at an International Statistical Institute (ISI)
World Statistics Congress, which took place at the Congress held in Dublin,
Ireland in August 2011.}
%
%
\maketitle

\begin{quote}\em
This perspective chapter briefly surveys:
(1)~past growth in the use of Bayesian methods in astrophysics; (2)~current
misconceptions about both frequentist and Bayesian statistical inference
that hinder wider adoption of Bayesian methods by astronomers; and
(3)~multilevel (hierarchical) Bayesian modeling as a major future direction
for research in Bayesian astrostatistics, exemplified in part by
presentations at the first ISI invited session on astrostatistics,
commemorated in this volume. It closes with an intentionally provocative
recommendation for astronomical survey data reporting, motivated by the
multilevel Bayesian perspective on modeling cosmic populations:  that
astronomers cease producing catalogs of estimated fluxes and other source
properties from surveys.  Instead, summaries of likelihood functions (or
marginal likelihood functions) for source properties should be reported (not
posterior probability density functions), including nontrivial summaries
(not simply upper limits) for candidate objects that do not pass traditional
detection thresholds.
\end{quote}

\noindent
This volume contains presentations from the first invited session on
astrostatistics to be held at an International Statistical Institute (ISI)
World Statistics Congress.  This session was a major milestone for
astrostatistics as an emerging cross-disciplinary research area.  It was the
first such session organized by the ISI Astrostatistics Committee, whose
formation in 2010 marked formal international recognition of the importance and
potential of astrostatistics by one of its information science parent
disciplines.  It was also a significant milestone for Bayesian
astrostatistics, as this research area was chosen as a (non-exclusive) focus
for the session.

As an early (and elder!) proponent of Bayesian methods in astronomy, I have
been asked to provide a ``perspective piece'' on the history and status of
Bayesian astrostatistics. I begin by briefly documenting the rapid rise in
use of the Bayesian approach by astrostatistics researchers over the past
two decades.  Next, I describe three misconceptions about both frequentist
and Bayesian methods that hinder wider adoption of the Bayesian approach
across the broader community of astronomer data analysts.  Then I
highlight the emerging role of multilevel (hierarchical) Bayesian modeling
in astrostatistics as a major future direction for research in Bayesian
astrostatistics.  I end with a provocative recommendation for survey data
reporting, motivated by the multilevel Bayesian perspective on modeling
cosmic populations.

\section{Looking back}
\label{sec:intro}

Bayesian ideas entered modern astronomical data analysis in the late 1970s,
when Steve Gull \& Geoff Daniell (1978, 1979) framed astronomical image deconvolution in
Bayesian terms.\footnote{Notably, Peter Sturrock (1973) earlier introduced
astronomers to the use of Bayesian probabilities for ``bookkeeping'' of
subjective beliefs about astrophysical hypotheses, but he did not discuss
statistical modeling of measurements per se.}
Motivated by Harold Jeffreys' Bayesian {\em Theory of Probability} (Jeffreys
1961), and Edwin Jaynes's introduction of Bayesian and information theory
methods into statistical mechanics and experimental physics (Jaynes 1959),
they addressed image estimation by writing down Bayes's theorem for the
posterior probability for candidate images, adopting an entropy-based prior
distribution for images.  They focused on finding a single ``best'' image
estimate based on the posterior:  the maximum entropy (MaxEnt) image
(maximizing the entropic prior probability subject to a somewhat ad hoc
likelihood-based constraint expressing goodness of fit to the data).  Such
estimates could also be found using frequentist penalized likelihood or
regularization approaches.  The Bayesian underpinnings of MaxEnt image
deconvolution thus seemed more of a curiosity than the mark of a major
methodological shift.

By about 1990, genuinely Bayesian data analysis---in the sense of reporting
Bayesian probabilities for statistical hypotheses, or samples from Bayesian
posterior distributions---began appearing in astronomy.  The Cambridge
MaxEnt group of astronomers and physicists, led by Gull and John
Skilling, began developing ``quantified MaxEnt'' methods to quantify
uncertainty in image deconvolution (and other inverse problems), rather than
merely reporting a single best-fit image.  On the statistics side, Brian
Ripley's group began using Gibbs sampling to sample from posterior
distributions for astronomical images based on Markov random field priors
(Ripley 1992).  My PhD thesis (defended in 1990) introduced parametric
Bayesian modeling of Poisson counting and point processes (including
processes with truncation or thinning, and measurement error) to high-energy
astrophysics (X-ray and gamma-ray astronomy) and to particle astrophysics
(neutrino astronomy). Bayesian methods were just beginning to be used for
parametric modeling of ground- and space-based cosmic microwave background
(CMB) data (e.g., Readhead \& Lawrence 1992).

It was in this context that the first session on Bayesian methods to be held
at an astronomy conference (to my knowledge) took place, at the first {\em
Statistical Challenges in Modern Astronomy} conference (SCMA~I), hosted by
statistician G.~Jogesh Babu and astronomer Eric Feigelson at Pennsylvania
State University in August 1991.  Bayesian methods were not only new, but
also controversial in astronomy at that time.  Of the 22 papers published in
the SCMA~I proceedings volume (Feigelson \& Babu 1992), only two were
devoted to Bayesian methods (Ripley 1992 and Loredo 1992a; see also the
unabridged version of the latter, Loredo 1992b).\footnote{A third paper
(Nousek 1992) had some Bayesian content but focused on frequentist
evaluation criteria, even for the one Bayesian procedure considered; these
three presentations, with discussion, comprised the Bayesian session.}  Both
papers had a strong pedagogical component (and a bit of polemic).  Of the
131 SCMA~I participants (about 60\% astronomers and 40\% statisticians),
only two were astronomers whose research prominently featured Bayesian
methods (Gull and me).

Twenty years later, the role of Bayesian methods in astrostatistics research
is dramatically different.  The 2008 Joint Statistical Meetings included two
sessions on astrostatistics predominantly devoted to Bayesian research. At
SCMA~V, held in June 2011, two sessions were devoted entirely to Bayesian
methods in astronomy:  ``Bayesian analysis across astronomy,'' with
eight papers and two commentaries, and ``Bayesian cosmology,'' including
three papers with individual commentaries. Overall, 14 of 32 invited
SCMA~V presentations (not counting commentaries) featured Bayesian methods,
and the focus was on calculations and results rather than on pedagogy and
polemic. About two months later, the ISI World Congress session on
astrostatistics commemorated in this volume was held; as already noted, its
focus was Bayesian astrostatistics.

On the face of it, these events seem to indicate that Bayesian methods are
not only no longer controversial, but are in fact now widely used, even favored
for some applications (most notably for parametric modeling in cosmology).  But
how representative are the conference presentations of broader astrostatistical
practice?

\begin{figure}[t]
\centerline{\includegraphics[width=.95\textwidth]{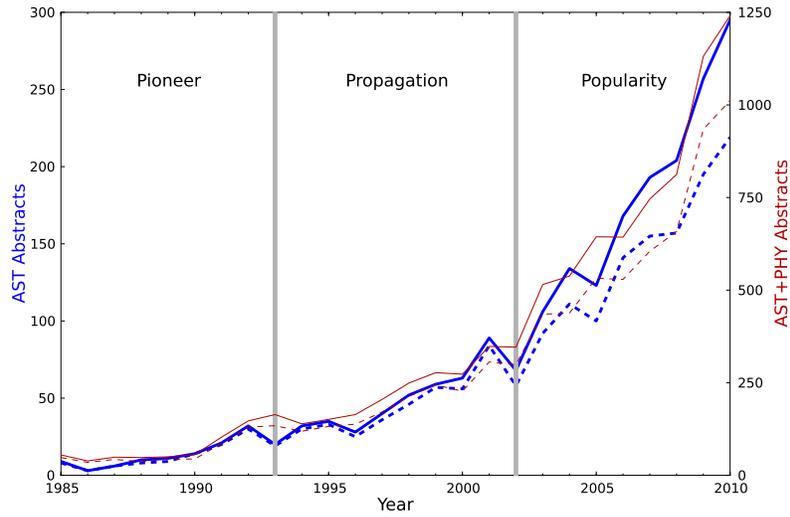}}
%
%
\caption{Simple bibliometrics measuring the growing use of Bayesian methods
in astronomy and physics, based on queries of the NASA ADS database in
October 2011. Thick
(blue) curves (against the left axis) are from queries of the astronomy
database; thin (red) curves (against the right axis) are from joint queries
of the astronomy and physics databases.  For each case the dashed lower
curve indicates the number of papers each year that include ``Bayes'' or
``Bayesian'' in the title or abstract.  The upper curve is based on the same
query, but also counting papers that use characteristically Bayesian
terminology in the abstract (e.g., the phrase ``posterior distribution'' or
the acronym ``MCMC''); it is meant to capture Bayesian usage in areas where
the methods are well-established, with the ``Bayesian'' appellation no
longer deemed necessary or notable.}
\label{fig:bib}
\end{figure}

Fig.~\ref{fig:bib} shows my amateur attempt at bibliometric measurement of
the growing adoption of Bayesian methods in both astronomy and physics,
based on queries of publication data in the NASA Astrophysics Data System
(ADS).  Publication counts indicate significant and rapidly growing use of
Bayesian methods in both astronomy and physics.\footnote{Roberto Trotta and
Martin Hendry have shown similar plots in various venues, helpfully noting
that the recent rate of growth apparent in Fig.~\ref{fig:bib} is much
greater than the rate of growth in the number of {\em all} publications;
i.e., not just the amount but also the prevalence of Bayesian work is rapidly
rising.}
Cursory examination of the publications reveals that Bayesian methods
are being developed across a wide range of astronomical subdisciplines.

It is tempting to conclude from the conference and bibliometric indicators
that Bayesian methods are now well-established and well-understood across
astronomy.  But the conference metrics reflect the role of Bayesian methods
in the astrostatistics research community, not in bread-and-butter
astronomical data analysis.  And as impressive as the trends in the
bibliometric metrics may be, the absolute numbers remain small in comparison
to all astronomy and physics publications, even limiting consideration to
data-based studies.  Although their impact is growing, Bayesian methods are
not yet in wide use by astronomers.

My interactions with colleagues indicate that significant misconceptions 
persist about fundamental aspects of both frequentist and Bayesian
statistical inference, clouding understanding of how these rival approaches
to data analysis differ and relate to one another.  I believe these
misconceptions play no small role in hindering broader adoption of Bayesian
methods in routine data analysis.  In the following section I highlight a
few misconceptions I frequently encounter.  I present them here as a
challenge to the Bayesian astrostatistics community; addressing them may
accelerate the penetration of sound Bayesian analysis into routine
astronomical data analysis.

\section{Misconceptions}
\label{sec:miscon}

For brevity, I will focus on just three important misconceptions I
repeatedly encounter about Bayesian and frequentist methods, posed as
(incorrect!) ``conceptual equations.''  They are:
\begin{itemize}
\item {\bf Variability $=$ Uncertainty}:  This is a misconception about
frequentist statistics that leads analysts to think that good frequentist
statistics is easier to do than it really is.
\item {\bf Bayesian computation $=$ Hard}:  This is a misconception about
Bayesian methods that leads analysts to think that implementing them
is harder than it really is, in particular, that it is harder than
implementing a frequentist analysis with comparable capability.
\item {\bf Bayesian = Frequentist $+$ Priors}: This is a misconception about
the role of prior probabilities in Bayesian methods that distracts analysts
from more essential distinguishing features of Bayesian inference.
\end{itemize}
I will elaborate on these faulty equations in turn.


\medskip
\noindent
{\bf Variability $=$ Uncertainty}: Frequentist statistics gets its name from
its reliance on the long-term frequency conception of probability:
frequentist probabilities describe the long-run variability of outcomes in
repeated experimentation.  Astronomers who work with data learn early in
their careers how to quantify the variability of data analysis procedures in
quite complicated settings using straightforward Monte Carlo methods that
simulate data.  How to get from quantification of {\em variability} in the
outcome of a procedure applied to an ensemble of simulated data, to a
meaningful and useful statement about the {\em uncertainty} in the conclusions
found by applying the procedure to the one actually observed data set, is a
subtle problem that has occupied the minds of statisticians for over a
century.  My impression is that many astronomers fail to recognize the
distinction between variability and uncertainty, and thus fail to appreciate
the achievements of frequentist statistics and their relevance to data
analysis practice in astronomy.  The result can be reliance on overly
simplistic ``home-brew'' analyses that at best may be suboptimal, but that
sometimes can be downright misleading.  A further consequence is a failure to
recognize fundamental differences between frequentist and Bayesian approaches
to quantification of uncertainty (e.g., that Bayesian probabilities for
hypotheses are not statements about variability of results in repeated
experiments).

To illustrate the issue, consider estimation of the parameters of a model
being fit to astronomical data, say, the parameters of a spectral model.  It
is often straightforward to find best-fit parameters with an optimization
algorithm, e.g., minimizing a $\chi^2$ measure of misfit, or maximizing a
likelihood function.  A harder but arguably more important task is
quantifying uncertainty in the parameters.  For abundant data, an asymptotic
Gaussian approximation may be valid, justifying use of the Hessian matrix
returned by many optimization codes to calculate an approximate
covariance matrix for defining confidence regions.  But when uncertainties are
significant and models are nonlinear, we must use more complicated
procedures to find accurate confidence regions.

Bootstrap resampling is a powerful framework statisticians use to develop
methods to accomplish this. There is a vast literature on applying the
bootstrap idea in various settings; much of it is devoted to the nontrivial
problem of devising algorithms that enable useful and accurate uncertainty
statements to be derived from simple bootstrap variability calculations.
Unfortunately, this literature is little-known in the astronomical
community, and too often astronomers misuse bootstrap ideas. The
variability-equals-uncertainty misconception appears to be at the root of
the problem.

\begin{figure}[t]
\centerline{\includegraphics[width=.5\textwidth]{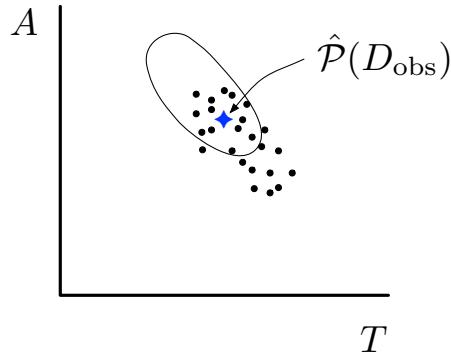}}
%
%
\caption{Illustration of the nontrivial relationship between variability of
an estimator, and uncertainty of an estimate as quantified by a frequentist
confidence region.  Shown is a two-dimensional parameter space with a best-fit
estimate to the observed data (blue 4-pointed star), best-fit estimates to
boostrapped data (black dots) showing variability of the estimator, and a
contour bounding a parametric bootstrap confidence region quantifying
uncertainty in the estimate.}
\label{fig:boot}
\end{figure}

As a cartoon example, suppose we have spectral data from a source that we
wish to fit with a simple thermal spectral model with two parameters, an
amplitude, $A$ (e.g., proportional to the source area and inversely
proportional to its distance squared), and a temperature, $T$ (determining
the shape of the spectrum as a function of energy or wavelength); we denote
the parameters jointly by $\params = (A,T)$. Fig.~\ref{fig:boot} depicts the
two-dimensional $(A,T)$ parameter space, with the best-fit parameters,
$\hat\params(\Dobs)$, indicated by the blue four-pointed star.  We can use
simulated data to find the variability of the estimator (i.e., of the
function $\hat\params(D)$ defined by the optimizer) were we to repeatedly
observe the source.  But how should we simulate data when we do not know the
true nature of the signal (and perhaps of noise and instrumental
distortions)?  And how should the variability of simulation results be used
to quantify the uncertainty in inferences based on the one observed dataset?

The underlying idea of the bootstrap is to use the observed dataset to
define the ensemble of hypothetical data to use in variability calculations,
and to find functions of the data (statistics) whose variability can be
simply used to quantify uncertainty (e.g., via moments or a histogram). 
Normally using the observed data to define the ensemble for simulations
would be cheating and would invalidate one's inferences; all frequentist
probabilities formally must be ``pre-observation'' calculations.  A major
achievement of the bootstrap literature is showing how to use the observed
data in a way that gives approximately valid results (hopefully with a rate
of convergence better than the $O(1/\sqrt{N})$ rate achieved by simple
Gaussian approximations, for sample size $N$).

One way to proceed is to use the full model we are fitting (both the signal model
and the instrument and noise model) to simulate data, with the parameters
fixed at $\hat\params(\Dobs)$ as a surrogate for the true values. 
Statisticians call this the {\em parametric bootstrap}; it was popularized
to astronomers in a well-known paper by Lampton, Margon \& Bowyer (1976,
hereafter LMB76; statisticians introduced the ``parametric bootstrap''
terminology later).  Alternatively, if some probabilistic aspects of the
model are not trusted (e.g., the error distribution is considered unrealistic),
an alternative approach is the {\em nonparametric bootstrap}, which
``recycles'' the observed data to generate simulated data (in some simple
cases, this may be done by sampling from the observed data with replacement
to generate each simulated data set).  Whichever approach we adopt, we will
generate a set of simulated data, $\{D_i\}$, to which we can apply our
fitting procedure to generate a set of best-fit parameter points
$\{\hat\params(D_i)\}$ that together quantify the variability of our
estimator.  The black dots in Fig.~\ref{fig:boot} show a scatterplot
or ``point cloud'' of such parameter estimates.

What does the point cloud tell us about the uncertainty we should associate
with the estimate from the observed data (the blue star)?  The practice
I have seen in too many papers is to interpret the points as samples from
a probability distribution in parameter space.\footnote{I am not providing
references to publications exhibiting the problem for diplomatic reasons and
for a more pragmatic and frustrating reason:  In the field where I have
repeatedly encountered the problem---analysis of infrared exoplanet transit
data---authors routinely fail to describe their analysis methods with
sufficient detail to know what was done, let alone to enable readers to
verify or duplicate the analysis. While there are clear signs of statistical
impropriety in many of the papers, I only know the details from personal
communications with exoplanet transit scientists.}
The cloud itself might be shown, with a contour enclosing a specified
fraction of the points offered as a joint confidence region. 
One-dimensional histograms may be used to find ``1$\sigma$'' (i.e., 68.3\%)
confidence regions for particular parameters of interest.  Such procedures
naively equate variability with uncertainty:  the uncertainty of the
estimate is identified with the variability of the estimator.  Regions created
this way are {\em wrong}, plainly and simply.  They will not cover the true
parameters with the claimed probability, and they are skewed in the wrong
direction.  This is apparent from the figure; the black points are skewed
{\em down} and to the {\em right} of the star indicating the parameter values
used to produce the simulated data; the parameters that produced the observed
data are thus likely to be {\em up} and to the {\em left} of the star.

In a correct parametric bootstrapping calculation (i.e., with a
trusted model), one can use simulated data to calibrate $\chi^2$ or
likelihood contours for defining confidence regions.  The procedure is not
very complicated, and produces regions like the one shown by the contour in
Fig.~\ref{fig:boot}, skewed in just the right way; LMB76 described the
construction.  But frequently investigators are attempting a {\em
non}parametric bootstrap, corresponding to a trusted signal model but an
untrusted noise model (often this is done without explicit justification, as
if nonparametric bootstrapping is the only type of bootstrapping).  In this
case devising a sound bootstrap confidence interval algorithm is not so
simple. Indeed, there is a large statistics literature on nonparametric
bootstrap confidence intervals, its size speaking to nontrivial challenges
in making the bootstrap idea work.  In particular, no simple procedure is
currently known for finding accurate joint confidence regions for multiple
parameters in nonlinear models using the nonparametric bootstrap; yet
results purporting to come from such a procedure appear in a number of
astronomical publications.  Strangely, the most cited reference on bootstrap
confidence regions in the astronomical literature is a book on numerical
methods, authored by astronomers, with an extremely brief and misleadingly
simplistic discussion of the nonparametric bootstrap (and no explicit
discussion of parametric bootstrapping).  Sadly, this is not the only area
where our community appears content disregarding a large body of relevant
statistics research (frequentist or otherwise).


What explains such disregard of relevant and nontrivial expertise? I am sure
there are multiple factors, but I suspect an important one is the
misconception that variability may be generically identified with
uncertainty. If one believes this, then doing (frequentist) statistics
appears to be a simple matter of simulating data to quantify the variability
of procedures.  For linear models with normally-distributed errors, the
identification is practically harmless; it is conceptually flawed but leads
to correct results by accident.  But more generally, determining how to use
variability to quantify uncertainty can be subtle and challenging. 
Statisticians have worked for a century to establish how to map variability
to uncertainty; when we seek frequentist quantifications of uncertainty in
nontrivial settings, we need to mine their expertise.

Why devote so much space to a misconception about frequentist statistics in
a commentary on Bayesian methods?  The variability-equals-uncertainty
misconception leads data analysts to think that frequentist statistics is
easier than it really is, in fact, to think that they already know what they
need to know about it.  If astronomers realize that sound statistical
practice is nontrivial and requires study, they may be more likely to study
Bayesian methods, and more likely to come to understand the
differences between the frequentist and Bayesian approaches.  Also, for the
example just described, the way some astronomers misuse the bootstrap is to
try to use it to define a probability distribution over parameter space. 
Frequentist statistics denies such a concept is meaningful, but it is
exactly what Bayesian methods aim to provide, e.g., with point clouds
produced via Markov chain Monte Carlo (MCMC) posterior sampling algorithms.
This brings us to the topic of Bayesian computation.


\medskip
\noindent
{\bf Bayesian computation $=$ Hard}:  A too-common (and largely unjustified)
complaint about Bayesian methods is that their computational implementation
is difficult, or more to the point, that Bayesian computation is harder than
frequentist computation. Analysts wanting a quick and useable result are
thus dissuaded from considering a Bayesian approach.  It is certainly true
that posterior sampling via MCMC---generating pseudo-random parameter values
distributed according to the posterior---is harder to do well than is
generating pseudo-random data sets from a best-fit model. (In fact, I would
argue that our community may not appreciate how hard it can be to do MCMC
well.)  But this is an apples-to-oranges comparison between methods
making very different types of approximations.  Fair comparisons, between
Bayesian and frequentist methods {\em with comparable capabilities and
approximations}, tell a different story.  Advocates of Bayesian methods need
to tell this story, and give this complaint a proper and public burial.

Consider first basic ``textbook'' problems that are analytically accessible.
Examples include estimating the mean of a normal distribution, estimating the
coefficients of a linear model (for data with additive Gaussian noise),
similar estimation when the noise variance is unknown (leading to Student's
$t$ distribution), estimating the intensity of a Poisson counting
process, etc..  In such problems, the Bayesian calculation is typically {\em
easier} than its frequentist counterpart, sometimes significantly so. 
Nowhere is this more dramatically demonstrated than in Jeffreys's classic
text, {\em Theory of Probability} (1961).  In chapter after chapter,
Jeffreys solves well-known statistics problems with arguments significantly
more straightforward and mathematics significantly more accessible than are
used in the corresponding frequentist treatments.  The analytical
tractability of such foundational problems is an aid to developing sound
statistical intuition, so this is not a trivial virtue of the Bayesian
approach.  (Of course, ease of use is no virtue at all if you reject an
approach---Bayesian, frequentist, or otherwise---on philosophical grounds.)

Turn now to problems of more realistic complexity, where approximate
numerical methods are necessary.  The main computational challenge for both
frequentist and Bayesian inference is multidimensional integration, over the
sample space for frequentist methods, and over parameter space for Bayesian
methods.  In high dimensions, both approaches tend to rely on Monte Carlo
methods for their computational implementation.  The most celebrated
approach to Bayesian computation is MCMC, which builds a multivariate
pseudo-random number generator producing dependent samples from a posterior
distribution.  Frequentist calculation instead uses Monte Carlo methods
to simulate data (i.e., draw samples from the sampling distribution). 
Building an MCMC posterior sampler, and properly analyzing its output, is
certainly more challenging than simulating data from a model, largely
because data are usually independently distributed; nonparametric bootstrap
resampling of data may sometimes be simpler still.  But the comparison is
not fair.  In typical frequentist Monte Carlo calculations, one simulates
data from the best-fit model (or a bootstrap surrogate), not from the true
model (or other plausible models).  The resulting frequentist quantities are
approximate, not just because of Monte Carlo error, but because the {\em
integrand} (or family of integrands) that would appear in an exact
frequentist calculation is being approximated (asymptotically).  In some
cases, an exact finite-sample frequentist formulation of the problem may not
even be known.  In contrast, MCMC posterior sampling makes no approximation of
integrands; results are approximate only due to Monte Carlo sampling error. No
large-sample asymptotic approximations need be invoked.

This point deserves amplification.  Although the main computational
challenge for frequentist statistics is integration over sample space,
there are additionally serious {\em theoretical} challenges for finite-sample
inference in many realistically complex settings.  Such challenges do
not typically arise for Bayesian inference.  These theoretical challenges,
and the analytical approximations that are adopted to address them,
are ignored in comparisons that pit simple Monte Carlo simulation of
data, or bootstrapping, against MCMC or other nontrivial Bayesian 
computation techniques.  Arguably, accurate finite-sample parametric
inference is often computationally {\em simpler} for Bayesian methods,
because an accurate frequentist approach is simply impossible, and an
approximate calculation can quantify only the rate of convergence of the
approximation, not the actual accuracy of the specific calculation being
performed.

If one is content with asymptotic approximations, the fairer comparison is
between asymptotic frequentist and asymptotic Bayesian methods.  At the
lowest order, asymptotic Bayesian computation using the Laplace
approximation is not significantly harder than asymptotic frequentist
calculation.  It uses the same basic numerical quantities---point estimates
and Hessian matrices from an optimizer---but in different ways.  It provides
users with nontrivial new capabilities, such as the ability to marginalize
over nuisance parameters, or to compare models using Bayes factors that
include an ``Ockham's razor'' penalty not present in frequentist
significance tests, and that enable straightforward comparison of rival
models that need not be nested (with one being a special case of the other).
And the results are sometimes accurate to one order higher (in $1/\sqrt{N}$)
than corresponding frequentist results (making them potentially competitive
with some bootstrap methods seeking similar asymptotic approximation
rates).

Some elaboration of these issues is in Loredo (1999), including references
to literature on Bayesian computation up to that time.  A more recent
discussion of this misconception about Bayesian methods (and other
misconceptions) is in an insightful essay by the Bayesian economist
Christopher Sims, winner of the 2011 Nobel Prize in economics (Sims 2010).

\medskip
\noindent
{\bf Bayesian = Frequentist $+$ Priors}:  I recently helped organize and
present two days of tutorial lectures on Bayesian computation, as a prelude
to the SCMA~V conference mentioned above.  As I was photocopying lecture
notes for the tutorials, a colleague walked into the copy room and had a
look at the table of contents for the tutorials. ``Why aren't all of the
talks about priors?'' he asked.  In response to my puzzled look, he
continued, ``Isn't that what Bayesian statistics is about, accounting for
prior probabilities?''

Bayesian statistics gets its name from Bayes's theorem, establishing that
the posterior probability for a hypothesis is proportional to the product of
its prior probability, and the probability for the data given the
hypothesis (i.e., the sampling distribution for the data).  The latter
factor is the likelihood function when considered as a function of the
hypotheses being considered (with the data fixed to their observed values). 
Frequentist methods that directly use the likelihood function, and their
least-squares cousins (e.g., $\chi^2$ minimization), are intuitively
appealing to astronomers and widely used.  On the face of it, Bayes's
theorem appears merely to add modulation by a prior to likelihood methods. 
By name, {\em Bayesian} statistics is evidently about using {\em Bayes's
theorem}, so it would seem it must be about how frequentist results should
be altered to account for prior probabilities.

It would be hard to overstate how wrong this conception of Bayesian
statistics is.

The name is unfortunate; Bayesian statistics uses {\em all} of probability
theory, not just Bayes's theorem, and not even primarily Bayes's theorem.
What most fundamentally distinguishes Bayesian calculations from
frequentist calculations is not modulation by priors, but the
key role of probability distributions {\em over parameter (hypothesis)
space} in the former, and the complete absence of such distributions
in the latter.  Via Bayes's theorem, a prior enables one to use the
likelihood function---describing a family of measures over the {\em sample}
space---to build the posterior distribution---a measure over the
{\em parameter} (hypothesis) space (where by ``measure'' I mean an additive
function over sets in the specified space).  This construction is just one
step in inference. Once it happens, the rest of probability theory kicks in,
enabling one to assess scientific arguments directly by calculating
probabilities quantifying the strengths of those arguments, rather than
indirectly, by having to devise a way that variability of a cleverly chosen
statistic across hypothetical data might quantify uncertainty across
possible choices of parameters or models for the actually observed data.

Perhaps the most important theorem for doing Bayesian calculations is the
{\em law of total probability} (LTP) that relates marginal probabilities to
sums of joint and conditional probabilities.  To display its role, suppose
we are analyzing some observed data, $\Dobs$, using a parametric model with
parameters $\theta$.  Let $M$ denote all the modeling
assumptions---definitions of the sample and parameter spaces, description of
the connection between the model and the data, and summaries of any relevant
prior information (parameter constraints or results of other measurements).
Now consider some of the common uses of  LTP in Bayesian analysis:
\begin{itemize}
\item Calculating the probability in a {\em credible region}, $R$, for $\theta$:
\begin{equation}
p(\theta \in R|\Dobs)
  = \int_R d\theta\, p(\theta \in R|\theta)\, p(\theta|\Dobs) \scond{M},
\label{HPDCR}
\end{equation}
where $p(\theta|\Dobs,M)$ is the posterior probability density
for $\theta$.
Here I have introduced a convenient shorthand due to John Skilling:
``$||M$'' indicates that $M$ is conditioning information
common to all displayed probabilities.
\item Calculating a {\em marginal posterior distribution} when a vector
parameter $\theta = (\psi,\eta)$ has both an interesting subset of
parameters, $\psi$, and uninteresting ``nuisance'' parameters, $\eta$.  The
uncertainty in $\psi$ (with the $\eta$ uncertainty fully propagated) is
quantified by the marginal density,
\begin{equation}
p(\psi |\Dobs) = \int d\eta\, p(\psi,\eta|\Dobs) \scond{M}.
\label{marg-pdf}
\end{equation}
\item Predicting future data, $D'$, with the posterior predictive distribution,
\begin{equation}
p(D'|\Dobs) = \int d\theta\, p(D'|\theta)\, p(\theta|\Dobs) \scond{M},
\label{post-predict}
\end{equation}
with the integration accounting for parameter uncertainty in the prediction.
\item Comparing rival parametric models $M_i$ (each with parameters $\theta_i$)
via posterior odds or Bayes factors, which requires computation of the
{\em marginal likelihood} for each model given by
\begin{equation}
p(\Dobs|M_i) = \int d\theta_i\, p(\theta_i|M_i)\, p(\Dobs|\theta_i,M_i)
  \scond{M_1\lor M_2\ldots}.
\label{marg-like}
\end{equation}
In words, this says that the likelihood for a model is the average of
the likelihood function for that model's parameters.
\end{itemize}
Arguably, if this approach to inference is to be named for a theorem,
``total probability inference'' would be a more appropriate appellation
than ``Bayesian statistics.''  It is probably too late to change the
name.  But it is not too late to change the emphasis.

In axiomatic developments of Bayesian inference, priors play no fundamental
role; rather, they {\em emerge} as a required ingredient when one seeks a
consistent or coherent calculus for the strengths of arguments that reason
from data to hypotheses.  Sometimes priors are eminently useful, as when one
wants to account for a positivity constraint on a physical parameter, or to
combine information from different experiments or observations.  Other times
they are frankly a nuisance, but alas still a necessity.

A physical analogy I find helpful for elucidating the role of priors in
Bayesian inference appeals to the distinction between intensive and
extensive quantities in thermodynamics.  Temperature is an intensive
property; in a volume of space it is meaningful to talk about the
temperature $T(x)$ at a {\em point} $x$, but not about the ``total
temperature'' of the {\em volume}; temperature does not add or integrate
across space.  In contrast, heat is an extensive property, an additive
property of volumes; in mathematical parlance, it may be described by a
measure (a mapping from regions, rather than points, to a real number).
Temperature and heat are related; the heat in a volume $V$ is given by
$Q = \int_V dx\, [\rho(x)c(x)] T(x)$, where $\rho(x)$ is the density and
$c(x)$ is the specific heat capacity.  The product $\rho c$ is extensive,
and serves to convert the intensive temperature to its extensive relative,
heat. In Bayesian inference, the prior plays an analogous role, not just
``modulating'' likelihood, but converting intensive likelihood to extensive
probability.  In thermodynamics, a region with a high temperature may have a
small amount of heat if its volume is small, or if, despite having a large
volume, the value of $\rho c$ is small.  In Bayesian statistics, a region of
parameter space with high likelihood may have a small probability if its
volume is small, or if the prior assigns low probability to the region.

This {\em accounting for volume in parameter space} is a key feature of
Bayesian methods.  What makes it possible is having a measure over parameter
space.  Priors are important, not so much as modulators of likelihoods, but
as converters from intensities (likelihoods) to measures (probabilities). 
With poetic license, one might say that frequentist statistics focuses on
the ``hottest'' (highest likelihood) hypotheses, while Bayesian inference
focuses on hypotheses with the most ``heat'' (probability).

\medskip
\noindent
{\bf Incommensurability}:
The growth in the use of Bayesian methods in recent decades has sometimes
been described as a ``revolution,'' presumably alluding to Thomas Kuhn's
concept of scientific revolutions (Kuhn 1970).  Although adoption of
Bayesian methods in many disciplines has been growing steadily and sometimes
dramatically, Bayesian methods have yet to completely or even
substantially replace frequentist methods in any broad discipline I am aware
of (although this has happened in some subdisciplines).
I doubt the pace and extent of change qualifies for a Kuhnian revolution.
Also, the Bayesian and frequentist approaches are not rival scientific
theories, but rather rival paradigms for a part of the scientific method
itself (how to build and assess arguments from data to scientific hypotheses).
Nevertheless, the competition between Bayesian and frequentist approaches to
inference does bear one hallmark of a Kuhnian revolution:
{\em incommensurability}.  I believe Bayesian-frequentist incommensurability
is not well-appreciated, and that it underlies multiple
misconceptions about the approaches.

Kuhn insisted that there could be no neutral or objective measure allowing
comparison of competing paradigms in a budding scientific revolution.  He
based this claim on several features he found common to scientific
revolutions, including the following:  (1)~Competing paradigms often adopt
different meanings for the same term or statement, making it very difficult
to effectively communicate across paradigms (a standard illustration is the
term ``mass,'' which takes on different meanings in Newtonian and
relativistic physics).  (2)~Competing paradigms adopt different
standards of evaluation; each paradigm typically ``works'' when judged by its
own standards, but the standards themselves are of limited use in comparing
across paradigms.


These aspects of Kuhnian incommensurability are evident in the frequentist
and Bayesian approaches to statistical inference.  (1)~The term
``probability'' takes different meanings in frequentist and Bayesian
approaches to uncertainty quantification, inviting misunderstanding when
comparing frequentist and Bayesian answers to a particular inference
problem.  (2)~Long-run performance is the gold standard for frequentist
statistics; frequentist methods aim for specified performance across repeated
experiments by construction, but make no probabilistic claims about the
result of application of a procedure to a particular observed dataset. 
Bayesian methods adopt more abstract standards, such as coherence or
internal consistency, that apply to inference for the case-at-hand, with no
fundamental role for frequency-based long-run performance.  
A frequentist method with good long-run performance can violate Bayesian
coherence or consistency requirements so strongly as to be obviously
unacceptable for inference in particular cases.\footnote{Efron (2003)
describes some such cases by saying the frequentist result can be {\em accurate}
but not {\em correct}.  Put another way, the performance claim is
{\em valid}, but the long-run performance can be {\em irrelevant} to the
case-at-hand, e.g., due to the existance of so-called recognizable subsets
in the sample space (see Loredo (1992) and Efron (2003) for elaboration
of this notion).  This is a further example of how nontrivial the relationship
between variability and uncertainty can be.}
On the other hand, Bayesian algorithms do not have guaranteed frequentist
performance; if it is of interest, it must be separately evaluated, and
priors may need adjustment to improve frequentist
performance.\footnote{There are theorems linking single-case Bayesian
probabilities and long-run performance in some general settings, e.g.,
establishing that, for fixed-dimension parametric inference, Bayesian
credible regions with probability $P$ have frequentist coverage close to $P$
(the rate of convergence is $O(1/\sqrt{N})$ for flat priors, and faster for
so-called reference priors).  But the theorems do not apply in some
interesting classes of problems, e.g., nonparametric problems.}



Kuhn considered rival paradigms to be so ``incommensurable'' that
``proponents of competing paradigms practice their trades in different
worlds.'' He argued that incommensurability is often so stark that for an
individual scientist to adopt a new paradigm requires a psychological shift
that could be termed a ``conversion experience.''  Following Kuhn,
philosophers of science have debated how extreme incommensurability really
is between rival paradigms, but the concept is widely considered important
for understanding significant changes in science.  In this context, it is
notable that statisticians, and scientists more generally, often adopt a
particular almost-religious terminology in frequentist vs.\ Bayesian
discussions: rather than a {\em method} being described as frequentist or
Bayesian, the {\em investigator} is so described.  This seems to me to be an
unfortunate tradition that should be abandoned.  Nevertheless, it does
highlight the fundamental incommensurability between these rival paradigms
for uncertainty quantification.  Advocates of one approach or the other (or
of a nuanced combination) need to more explicitly note and discuss this
incommensurability, especially with nonexperts seeking to choose between
approaches.



The fact that both paradigms remain in broad use suggests that ideas from
both approaches may be relevant to inference; perhaps they are each suited
to addressing different types of scientific questions.  For example,
my discussion of misconceptions has been largely from the perspective of
parametric modeling (parameter estimation and model comparison).  
{\em Non}parametric inference raises more subtle issues regarding both
computation and the role of long-term performance in Bayesian inference; see
Bayarri \& Berger (2004) and Sims (2010) for insightful discussions of some
of these issues.  Also, Bayesian model checking (assessing the adequacy of a
parametric model without an explicitly specified class of alternatives)
typically invokes criteria based on predictive frequencies (Bayarri \&
Berger 2004; Gelman et al.\ 2004; Little 2006).  A virtue of the
Bayesian approach is that one may predict or estimate frequencies when they
are deemed relevant; explicitly distinguishing probability (as degree of
strength of an argument) from frequency (in finite or hypothetical infinite
samples) enables this.  This suggests some kind of unification of approaches
may be easier to achieve from the Bayesian direction.  This is a worthwhile
task for research; see Loredo (2012a) for a brief overview of some recent work
on the Bayesian/frequentist interface.

No one would claim that the Bayesian approach is a data analysis panacea,
providing the best way to address all data analysis questions.
But among astronomers outside of the community of astrostatistics
researchers, Bayesian methods are significantly underutilized.  Clearing up
misconceptions should go a long way toward helping astronomers appreciate
what both frequentist and Bayesian methods have to offer for both routine
and research-level data analysis tasks.

\section{Looking forward}
\label{sec:forward}

Having looked at the past growth of interest in Bayesian methods and present
misconceptions, I will now turn to the future.  As inspiration, I cite Mike
West's commentary on my SCMA~I paper (West 1992).  In his closing remarks he
pointed to an especially promising direction for future Bayesian work in
astrostatistics:

\begin{quotation}
On possible future directions, it is clear that Bayesian developments during
recent years have much to offer---I would identify prior modeling
developments in {\em hierarchical} models as particularly noteworthy.
Applications of such models have grown tremendously in biomedical and
social sciences, but this has yet to be paralleled in the physical
sciences.  Investigations involving repeat experimentation on
similar, related systems provide the archetype logical structure for
hierarchical modeling\ldots There are clear opportunities for exploitation
of these (and other) developments by astronomical investigators\ldots.
\end{quotation}

\noindent
However clear the opportunities may have appeared to West, for over a decade
following SCMA~I, few astronomers pursued hierarchical Bayesian modeling.  A
particularly promising application area is modeling of populations of
astronomical sources, where hierarchical models can naturally account for
measurement error, selection effects, and ``scatter'' of properties across a
population.  I discussed this at some length at SCMA~IV in 2006 (Loredo
2007), but even as of that time there was relatively little work in
astronomy using hierarchical Bayesian methods, and for the most part only
the simplest such models were used.

The last few years mark a change point in this respect, and evidence of the
change is apparent in the contributions to the summer 2011 Bayesian sessions
at both SCMA~V and the ISI World Congress. Several presentations in both
forums described recent and ongoing research developing sophisticated
hierarchical models for complex astronomical data.  Other papers raised
issues that may be addressed with hierarchical models.  Together, these
papers point to hierarchical Bayesian modeling as an important emerging
research direction for astrostatistics.

To illustrate the notion of a hierarchical model---also known as a {\em
multilevel model} (MLM)---we start with a simple parametric density
estimation problem, and then promote it to a MLM by adding measurement
error.

\begin{figure}[t]
\centerline{\includegraphics[width=\textwidth]{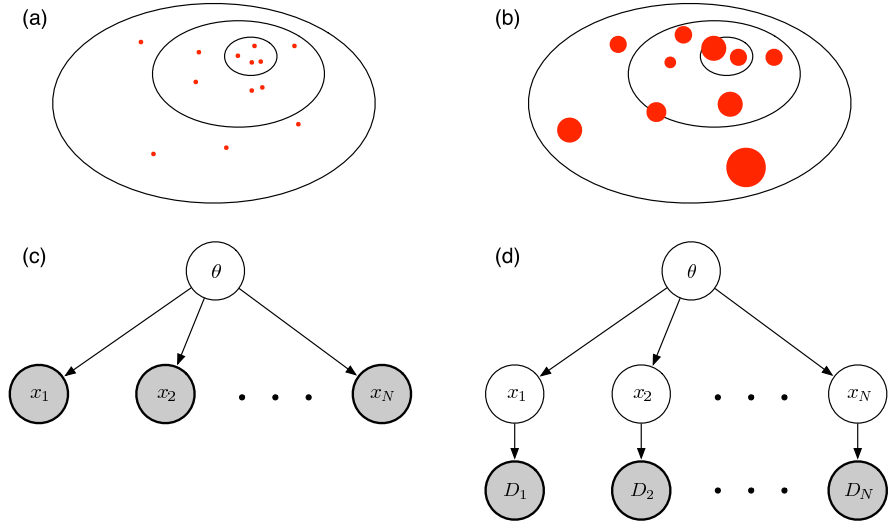}}
\caption{Illustration of multilevel model approach to handling
measurement error.  (a) and (b) (top row): 
Measurements of a two-dimensional observable and its probability
distribution (contours); in (a) the measurements are precise (points);
in (b) they are noisy (filled circles depict uncertainties).
(c) and (d):  Graphical models corresponding to Bayesian
estimation of the density in (a) and (b), respectively.}
\label{fig:MLMs1}
\end{figure}

Suppose we would like to estimate parameters $\theta$ defining
a probability density function $f(x;\theta)$ for an observable $x$.
A concrete example might be estimation of a galaxy luminosity function,
where $x$ would be two-dimensional, $x=(L,z)$ for luminosity $L$ and
redshift $z$, and $f(x;\theta)$ would be the normalized luminosity 
function (i.e., a probability density rather than a galaxy number density).
Consider first the case where we have a set of precise measurements
of the observables, $\{x_i\}$ (and no selection effects).  Panel~(a) in
Fig.~\ref{fig:MLMs1} depicts this simple setting.  The likelihood function
for $\theta$ is $\like(\theta) \equiv p(\{x_i\}|\theta,M) = \prod_i
f(x_i;\theta)$. Bayesian estimation of $\theta$ requires a prior density,
$\pi(\theta)$, leading to a posterior density
$p(\theta|\{x_i\}, M) \propto \pi(\theta)\like(\theta)$.

An alternative way to write Bayes's theorem expresses the posterior
in terms of the joint distribution for parameters and data:
\begin{equation}
p(\theta|\{x_i\}, M) = \frac{p(\theta,\{x_i\}| M)}{p(\{x_i\}|M)}.
\label{BT-joint}
\end{equation}
This ``probability for everything'' version of Bayes's theorem changes
the process of modeling from separate specification of a prior and likelihood,
to specification of the joint distribution for everything; this
proves helpful for building models with complex dependencies.
Panel~(c) depicts the dependencies in the joint distribution with a
graph---a collection of nodes connected by edges---where each node
represents a probability distribution for the indicated variable, and the
directed edges indicate dependences between variables.  Shaded nodes
indicate variables whose values are known (here, the data); we may
manipulate the joint to condition on these quantities.  The graph
structure visually displays how the joint distribution may be factored as a
sequence of independent and conditional distributions: the $\theta$ node
represents the prior, and the $x_i$ nodes represent
$f(x_i;\theta)=p(x_i|\theta,M)$ factors, dependent on $\theta$ but independent
of other $x_i$ values when $\theta$ is given (i.e., conditionally
independent).  The joint distribution is thus
$p(\theta, \{x_i\}|M) = \pi(\theta) \prod_i f(x_i;\theta)$.  In a sense, the
most important edges in the graph are the {\em missing} edges; they indicate
independence that makes factors simpler than they might otherwise be.

Now suppose that, instead of precise $x_i$ measurements, for each
observation we get noisy data, $D_i$, producing a measurement 
likelihood function $\ell_i(x_i) \equiv p(D_i|x_i,M)$ describing
the uncertainties in $x_i$ (we might summarize it with the mean
and standard deviation of a Gaussian).  Panel (b) depicts
the situation; instead of points in $x$ space, we now have
likelihood functions (depicted as ``1$\sigma$'' error circles).
Panel (d) shows a graph describing this measurement error problem,
which adds a $\{D_i\}$ level to the previous graph; we now have a
multilevel model.\footnote{The convention is to reserve the term for models
with three or more levels of nodes, i.e., two or more levels of edges, or
two or more levels of nodes for {\em uncertain variables} (i.e., unshaded
nodes).  The model depicted in panel~(d) would be called a two-level model.}
The $x_i$ nodes are now unshaded; they are no longer known, and have become
{\em latent parameters}.  From the graph we can read off the form of the
joint distribution:
\begin{equation}
p(\theta,\{x_i\},\{D_i\}|M) =
  \pi(\theta) \prod_i f(x_i;\theta) \ell_i(x_i).
\label{p-txd}
\end{equation}
From this joint distribution we can make inferences about any quantity of
interest.  To estimate $\theta$, we use the joint to calculate
$p(\theta,\{x_i\}|\{D_i\},M)$ (i.e., we condition on the known data using
Bayes's theorem), and then we marginalize over all of the latent $x_i$
variables.  We can estimate all the $x_i$ values jointly by instead
marginalizing over $\theta$.  Note that this produces a joint marginal
distribution for $\{x_i\}$ that is {\em not} a product of independent factors;
although the $x_i$ values are conditionally independent given $\theta$, they
are {\em marginally dependent}.  If we do not know $\theta$, each $x_i$ tells
us something about all the others through what it tells us about $\theta$.
Statisticians use the phrase ``borrowing strength'' to describe this effect,
from John Tukey's evocative description of ``mustering and borrowing
strength'' from related data in multiple stages of data analysis (see Loredo
and Hendry 2010 for a tutorial discussion of this effect and the related
concept of shrinkage estimators).  Note the prominent role of LTP in inference
with MLMs, where inference at one level requires marginalization over unknowns
at other levels.

\begin{figure}[t]
\centerline{\includegraphics[width=\textwidth]{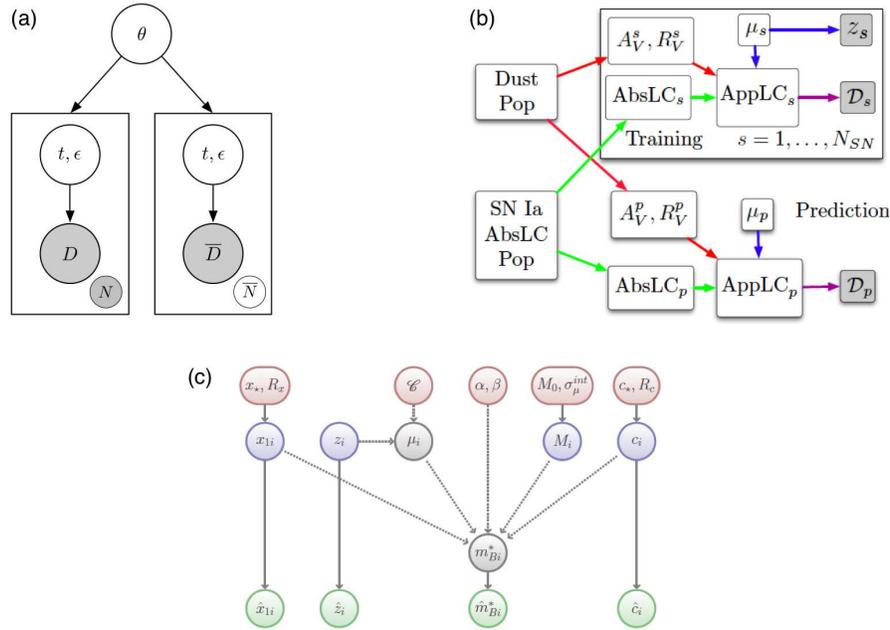}}
\caption{Graphs describing multilevel models used in astronomy, as
described in the text.}
\label{fig:MLMs2}
\end{figure}

The few Bayesian MLMs used by astronomers through the 1990s and early 2000s
did not go much beyond this simplest hierarchical structure.  For example,
unbeknownst to West, at the time of his writing my thesis work had already
developed a MLM for analyzing the arrival times and energies of neutrinos
detected from SN~1987A; the multilevel structure was needed to handle
measurement error in the energies (an expanded version of this work appears
in Loredo \& Lamb 2002).  Panel~(a) of Fig.~\ref{fig:MLMs2} shows a graph
describing the model.  The rectangles are ``plates'' indicating
substructures that are repeated; the integer variable in the corner
indicates the number of repeats.  There are two plates because neutrino
detectors have a limited (and energy-dependent) detection efficiency.  The
plate with a known repeat count, $N$, corresponds to the $N$ detected
neutrinos with times $t$ and energies $\epsilon$; the plate with an unknown
repeat count, $\overline N$, corresponds to undetected neutrinos, which must
be considered in order to constrain the total signal rate; $\overline D$
denotes the nondetection data, i.e., reports of zero events in time
intervals between detections.

Other problems tackled by astronomers with two-level MLMs include: 
modeling of number-size distributions (``$\log N$--$\log S$'' or ``number
counts'') of gamma-ray bursts and trans-Neptunian objects (e.g., Loredo \&
Wasserman 1998; Petit et al.\ 2008); performing linear regression with
measurement error along both axes, e.g., for correlating quasar hardness and
luminosity (Kelly 2007; see his contribution in Feigelson \& Babu 2012 for an
introduction to MLMs for measurement error) and for correlating
galaxy cluster richness and mass (Andreon \& Hurn 2010);
accounting for Eddington and
Malmquist biases in cosmology (Loredo \& Hendry 2010); statistical
assessment of directional coincidences with gamma-ray bursts (Luo, Loredo \&
Wasserman 1996; Graziani \& Lamb 1996) and cross-matching catalogs produced
by large star and galaxy surveys (Budav\'ari \& Szalay 2008; see Loredo
2012b for discussion of the underlying MLM); and handling multivariate
measurement error when estimating stellar velocity distributions from proper
motion survey data (Bovy, Hogg, \& Roweis 2011).

Beginning around 2000, interdisciplinary teams of astronomers and information
scientists began developing significantly more sophisticated MLMs for
astronomical data.  The most advanced such work has come from a
collaboration of astronomers and statisticians associated with the
Harvard-Smithsonian Center for Astrophysics (CfA).  Much of their work has
been motivated by analysis of data from the {\em Chandra} X-ray observatory
satellite, whose science operations are managed by CfA.  van~Dyk et 
al.\ (2001) developed a many-level MLM for fitting {\em Chandra} X-ray spectral
data; a host of latent parameters enable accurate accounting for uncertain
backgrounds and instrumental effects such as pulse pile-up.  Esch et
al.\ (2004) developed a Bayesian image reconstruction algorithm for {\em
Chandra} imaging data that uses a multiscale hierarchical prior to build
spatially-adaptive smoothing into image estimation and uncertainty
quantification.  van~Dyk et al.\ (2009) showed how to analyze stellar
cluster color-magnitude diagrams (CMDs) using finite mixture models (FMMs)
to account for contamination of the data from stars not lying in the
targeted cluster.  In FMMs, categorical class membership variables appear as
latent parameters; the mixture model effectively averages over many possible
graphs (corresponding to different partitions of the data into classes;
such averaging over partitions also appears in the Bayesian cross-matching
framework of Luo, Loredo \& Wasserman 1996).
FMMs have long been used to handle outliers and contamination in Bayesian
regression and density estimation.  This work showed how to implement it with
computationally expensive models and informative class membership priors.
In the area of time series, the astronomer-engineer collaboration of
Dobigeon, Tourneret, \& Scargle (2007) developed a three-level MLM to tackle
joint segmentation of astronomical arrival time series (a multivariate
extension of Scargle's well-known Bayesian Blocks algorithm).

Cosmology is a natural arena for multilevel modeling, because of the indirect
link between theory and observables.  For example, in modeling both the
cosmic microwave background (CMB) and the large scale structure (LSS) of
the galaxy distribution, theory does not predict a specific temperature map
or set of galaxy locations (these depend on unknowable initial conditions),
but instead predicts statistical quantities, such as angular or spatial
power spectra.  Modeling observables given theoretical parameters typically
requires introducing these quantities as latent parameters.  In Loredo
(1995) I described a highly simplified hierarchical treatment of CMB data,
with noisy CMB temperature difference time series data at the lowest level,
$l=2$ spherical harmonic coefficients in the middle, and a single physical
parameter of interest, the cosmological quadrupole moment $Q$, at the top. 
While a useful illustration of the MLM approach, I noted there were enormous
computational challenges facing a more realistic implementation.  It took a
decade for such an implementation to be developed, in the pioneering work
of Wandelt et al.\ (2004).  And only recently have explicit hierarchical
models been implemented for LSS modeling (e.g., Kitaura \& En\ss lin 2008).

This brings us to the present.  Contributions in this volume and in the
SCMA~V proceedings (Feigelson \& Babu 2012) document burgeoning interest in
Bayesian MLMs among astrostatisticians.  I discuss the role of MLMs in the
SCMA~V contributions elsewhere (Loredo 2012c).  Four of the contributions in
the present volume rely on MLMs.  Two address complex problems and help mark
an era of new complexity and sophistication in astrophysical MLMs.  Two
highlight the potential of basic MLMs for addressing common astronomical data
analysis problems that defy accurate analysis with conventional methods taught
to astronomers.  I will highlight the MLM aspects of these contributions in
turn.

Some of the most impressive new Bayesian multilevel modeling in astronomy
addresses the analysis of measurements of multicolor light curves
(brightness vs.\ time in various wavebands) from Type~Ia supernovae
(SNe~Ia). In the late 1990s, astronomers discovered that these enormous
stellar thermonuclear exoplosions are ``standardizable candles;'' the {\em
shapes} of their light curves are strongly correlated with their {\em
luminosities} (the intrinsic amplitudes of the light curves).  This
enables use of SNe~Ia to measure cosmic distances (via a generalized
inverse-square law) and to trace the history of cosmic expansion.  The 2011
Nobel Prize in physics went to three astronomers who used this capability
to show that the expansion rate of the universe is {\em growing} with time
(``cosmic acceleration''), indicating the presence of ``dark energy'' that
somehow prevents the deceleration one would expect from gravitational
attraction.

A high-water mark in astronomical Bayesian multilevel modeling
was set by Mandel et al.\ (2011), who address the problem of estimating
supernova luminosities from light curve data.  Their model has three levels,
complex connections between latent variables (some of them random {\em
functions}---light curves---rather than scalars), and jointly describes
three different types of data (light curves, spectra, and host galaxy
spectroscopic redshifts). Panel~(b) of Fig.~\ref{fig:MLMs2} shows the graph
for their MLM (the reader will have to consult Mandel et al. (2011), or
Mandel's more tutorial overview in Feigelson \& Babu (2012), for a
description of the variables and the model).  In this volume, March et
al.\ tackle a subsequent SNe~Ia problem: how to use the output of light
curve models to estimate cosmological parameters, including the density of
dark energy, and an ``equation of state'' parameter aiming to capture how
the dark energy density may be evolving.  Their framework can fuse
information from SN~Ia with information from other sources, such as the
power spectrum of CMB fluctuations, and characterization of the baryon
acoustic oscillation (BAO) seen in the large scale spatial distribution of
galaxies. Panel~(c) of Fig.~\ref{fig:MLMs2} shows the graph for their model,
also impressive for its complexity (see their contribution in this volume
for an explanation of this graph).  I display the graphs (without much
explanation) to show how much emerging MLM research in astronomy is
leapfrogging the simple models of the recent past, exemplified by Panel~(a)
in the figure.  Section~3.4 of Andreon's tutorial in this volume
discusses essentially the same model.

Equally impressive is Wandelt's contribution, describing a hierarchical
Bayes approach (albeit without explicit MLM language) for reconstructing the
galaxy density field from noisy photometric redshift data measuring the
line-of-sight velocities of galaxies (which includes a component from cosmic
expansion and a ``peculiar velocity'' component from gravitational
attraction of nearby galaxies and dark matter).  His team's framework
(described in detail by Jasche \& Wandelt 2011) includes nonparametric
estimation of the density field at an upper level, which adaptively
influences estimation of galaxy distances and peculiar velocities at a lower
level, borrowing strength in the manner described above.  The nonparametric
upper level, and the size of the calculation, make this work groundbreaking.

The contributions by Andreon and by Kunz et al.\ describe simpler MLMs, but
with potentially broad applicability.  Andreon's galaxy cluster mass
prediction example describes regression with
measurement error (fitting lines and curves using data with errors in both
the abscissa and ordinate) using a basic MLM.  Such problems are common in astronomy.  Kelly (2007; see
also Kelly's more introductory treatment in Feigelson \& Babu 2012) provided
a more formal account of  the Bayesian treatment of such problems, drawing
on the statistics literature on measurement error problems (e.g., Carroll et
al.\ 2006).  The complementary emphasis of Andreon's account is on how
straightforward---indeed, almost automatic---implementation can be using
modern software packages such as BUGS and
JAGS,\footnote{\url{http://www.mrc-bsu.cam.ac.uk/bugs/}}
a point also discussed (a bit more critically) by Carroll et al.\ (2006).
The contribution by Kunz et al.\ further develops the Bayesian estimation
applied to multiple species (BEAMS) framework first described by Kunz et al.\
(2007).  BEAMS aims to improve parameter estimation in nonlinear regression
when the data may come from different types of sources (``species'' or
classes), with different error or population distributions, but with
uncertainty in the type of each datum.  The classification labels for the data
become discrete latent parameters in a MLM; marginalizing over them (and
possibly estimating parameters in the various error distributions) can greatly
improve inferences. They apply the approach to estimating cosmological
parameters using SNe~Ia data, and show that accounting for uncertainty in
supernova classification has the potential to significantly improve the
precision and accuracy of estimates. In the context of this volume, one cannot
help but wonder what would come of integrating something like BEAMS into the
MLM framework of March et al..

In Section~2 I noted how the ``variability $=$ uncertainty'' misconception
leads many astronomers to ignore relevant frequentist statistics literature;
it gives the analyst the impression that the ability to quantify variability
is all that is needed to devise a sound frequentist procedure.  There is
a similar danger for Bayesian inference.  Once one has specified a model
for the data (embodied in the likelihood function), Bayesian inference
appears to be automatic in principle; one just follows the rules of
probability theory to calculate probabilities for hypotheses of interest
(after assigning priors).  But despite the apparent simplicity of the sum
and product rules, probability theory can exhibit a sophisticated subtlety,
with apparently innocuous assumptions and calculations sometimes producing
surprising results.  The huge literature on Bayesian methods is more than
mere crank turning; it amasses significant experience with this sophisticated
machinery that astronomers should mine to guide development of new Bayesian
methods, or to refine existing ones.  For this non-statistician reader, it is
the simpler of the contributions in this volume---those by Andreon and by
Kunz et al.---that point to opportunities for ``borrowing and mustering of
strength'' from earlier work by statisticians.

Andreon's ``de-TeXing'' approach to Bayesian modeling (i.e., transcribing
model equations to BUGS or JAGS code, and assigning simple default priors)
is both appealing and relatively safe for simple models, such as standard
regression (parametric curve fitting with errors only in the ordinate) or
density estimation from precise point measurements in low dimensions.
But the implications of modeling assumptions become increasingly
subtle as model complexity grows, particularly when one starts increasing
the number of levels or the dimensions of model components (or both).
Nontrivial computational challenges may also ensue.  This
makes me wary of emphasizing an automated style of Bayesian modeling. 
Let us focus here on some MLM subtlety that can complicate matters.

Information gain from the data tends to
weaken as one considers parameters at increasingly high levels
in a multilevel model (Goel \& DeGroot 1981).  On the one hand, if one is
interested in quantities at lower levels, this weakens dependence on
assumptions made at high levels.  On the other hand, if one is interested in
high-level quantities, sensitivity to the prior becomes an issue.  The
weakened impact of data on high levels has the effect that improper (i.e.,
non-normalized) priors that are safe to use in simple models (because the
likelihood makes the posterior proper) can be dangerous in MLMs, producing
improper posteriors; proper but ``vague'' default priors may hide the
problem without truly ameliorating
it (Hadjicostas \& Berry 1999; Gelman 2006).  Paradoxically, in some
settings one may need to assign very informative upper-level priors to allow
lower level distributions to adapt to the data (see Esch et al.\ 2004 for an
astronomical example).  Andreon's closing recommendations regarding sensitivity
analysis hint at some of these issues; his research (e.g.,
Andreon \& Hurt 2010) provides concrete examples of such analysis, which is
more important for MLMs than for simpler models.

In addition, the impact of the graph structure on a
model's predictive ability becomes less intuitively accessible as complexity
grows, making predictive tests of MLMs important, but also nontrivial.
Simple posterior predictive tests (such as described by Andreon in his
SN~Ia example) may be useful, but can often be insensitive to significant
model-data discrepancies (Sinharay \& Stern 2003; Gelman et al.\ 2004;
Bayarri \& Castellanos 2007).  An exemplary feature of the SNe~Ia MLM work
of Mandel et al.\ is the use of out-of-sample predictive checks, implemented
via a frequentist cross-validation procedure, to quantitatively assess the
adequacy of various aspects of the model (notably, Mandel audited
graduate-level statistics courses to learn the ins and outs of MLMs for this
work, comprising his PhD thesis).

In the context of nonlinear regression
with measurement error---Andreon's topic---Carroll et al.\ (2006) provides a
useful entry point to both Bayesian and frequentist literature, incidentally
also describing a number of frequentist approaches to such problems that
would be more fair competitors to Bayesian MLMs than ``the usual chi-squared
approach,'' Andreon's sadly accurate phrase for a naive weighted least
squares method shown by March et al.\ to produce inaccurate results when
measurement errors are significant. The variety of sound frequentist
approaches for handling measurement error belies Andreon's subsequent claim
that ``non-Bayesian methods [are] obliged to discard part of the available
information in order to reach the finishing line.''  Nevertheless, the
straightforwardness and flexibility of the Bayesian approach, and the ease of
interpretation of both the computations and the results, is leading 
astrostatisticians to share Andreon's enthusiasm for Bayesian MLM 
approaches to such problems.

The problem addressed by the BEAMS framework---essentially the problem of
data contamination, or mixed data---is not unique to astronomy, and there is
significant statistics literature addressing similar problems with lessons
to offer astronomers.  BEAMS is a form of Bayesian regression using FMM
error distributions.  Statisticians first developed such models a few
decades ago, to treat outliers (points evidently not obeying the assumed
error distribution) using simple two-component mixtures (e.g., normal
distributions with two different variances).  More sophisticated versions
have since been developed for diverse applications.  An astronomical example
building on some of this expertise is the finite mixture modeling of stellar
populations by van~Dyk et al.\ (2009), mentioned above.  The most immediate
lessons astronomers may draw from this literature are probably
computational; for example, algorithms using data augmentation (which
involves a kind of guided, iterative Monte Carlo sampling of the class
labels and weights) may be more effective for implementing BEAMS than the
weight-stepping approach described by Kunz et al.\ in this volume.


\section{Provocation}
\label{sec:provoke}

I will close this commentary with a provocative recommendation I have
offered at meetings since 2005 but not yet in print, born of my
experience using multilevel models for astronomical populations.  It is that
astronomers cease producing catalogs of estimated fluxes and other source
properties from surveys.  This warrants explanation and elaboration.


As noted above, a consequence of the hierarchical structure of MLMs is that
the values of latent parameters at low levels cannot be estimated
independently of each other.  In a survey context, this means that the flux
(and potentially other properties) of a source cannot be accurately or
optimally estimated considering only the data for that source.  This
may initially seem surprising, but at some level astronomers already
know this to be true.  We know---from Eddington, Malmquist, and Lutz \&
Kelker---that simple estimates of source properties will be misleading if we
do not take into account information besides the measurements and selection
effects; we also must specify the population distribution of the property.
The standard Malmquist and Lutz-Kelker corrections adopt an a priori fixed
(e.g., spatially homogeneous) population distribution, and produce a
corrected estimate independently for each object.  What the fully Bayesian MLM
approach adds to the picture is the ability to handle uncertainty in the
population distribution.  After all, a prime reason for performing surveys is
to learn about populations.  When the population distribution is not
well-known a priori, each source property measurement bears on estimation of
the population distribution, and thus indirectly, each measurement bears on
the estimation of the properties of every other source, via a kind of adaptive
bias correction (Loredo \& Hendry 2010).\footnote{It is worth pointing out
that this is not a uniquely Bayesian insight. Eddington, Malmquist, and Lutz
\& Kelker used frequentist arguments to justify their corrections; Eddington
even offered adaptive corrections.  The large and influential statistics
literature on shrinkage estimators leads to similar conclusions; see Loredo
(2007) for further discussion and references.}
This is Tukey's ``mustering and borrowing of strength'' at work again.

To enable this mustering and borrowing, we have to stop thinking of a
catalog entry as providing all the information needed to infer a particular
source's properties (even in the absence of auxiliary information from
outside a particular survey).  Such a complete summary of information is
provided by the marginal posterior distribution for that source, which
depends on the data from {\em all} sources---and on population-level
modeling assumptions.  However, in the MLM structure (e.g., panel (d) of
Fig.~\ref{fig:MLMs1}), the {\em likelihood function} for the properties of a
particular source may be independent of information about other sources. 
The simplest output of a survey that would enable accurate and optimal
subsequent analysis is thus a {\em catalog of likelihood functions} (or
possibly marginal likelihood functions when there are uncertain
survey-specific backgrounds or other ``nuisance'' effects the surveyor must
account for).

For a well-measured source, the likelihood function may be well-approximated
by a Gaussian that can be easily summarized with a mean and standard
deviation.  But these should not be presented as point estimates and
uncertainties.\footnote{I am tempted to recommend that, even in this regime,
the likelihood summary be chosen so as to deter misuse as an estimate, say
by tabulating the $+1\sigma$ and $-2\sigma$ points rather than means and
standard deviations. I am only partly facetious about this!}
For sources near the ``detection limit,'' more complicated summaries may be
justified. Counterpart surveys should cease reporting upper limits when a
known source is not securely detected; instead they should report a more
informative non-Gaussian likelihood summary.  Discovery surveys (aiming to
detect new sources rather than counterparts) could potentially devise
likelihood summaries that communicate information about sources with fluxes
{\em below} a nominal detection limit, and about uncertain source
multiplicty in crowded fields.  Recent work on maximum-likelihood fitting of
``pixel histograms'' (also known as ``probability of deflection'' or $P(D)$
distributions), which contain information about undetected sources, hints at
the science such summaries might enable in a MLM setting (e.g., Patanchon et
al.\ 2009).

In this approach to survey reporting, the notion of a detection
limit as a decision boundary identifying sources disappears.  In its place
there will be decision boundaries, driven by both computational and
scientific considerations, that determine what type of likelihood summary is
associated with each possible candidate source location.

Coming at this issue from another direction, Hogg \& Lang (2011) have
recently made similar suggestions, including some specific ideas for
how likelihoods may be summarized.  Multilevel models provide a principled
framework, both for motivating such a thoroughgoing revision of current
practice, and for guiding its detailed development.  Perhaps by
the 2015 ISI World Congress in Brazil we will hear reports of analyses
of the first survey catalogs providing such more optimal, MLM-ready
summaries.

But even in the absence of so revolutionary a development, I think one can
place high odds in favor of a bet that Bayesian multilevel modeling will
become increasingly prevalent (and well-understood) in forthcoming
astrostatistics research.  Whether Bayesian methods (multilevel and
otherwise) will start flourishing {\em outside} the astrostatistics research
community is another matter, dependent on how effectively astrostatisticians
can rise to the challenge of correcting misconceptions about both
frequentist and Bayesian statistics, such as those outlined above.  The
abundance of young astronomers with enthusiasm for astrostatistics makes
me optimistic.


%
\begin{acknowledgement}
I gratefully acknowledge NSF and NASA for support of current research
underlying this commentary, via grants AST-0908439, NNX09AK60G and
NNX09AD03G.  I thank Martin Weinberg for helpful discussions on information
propagation within multilevel models.  Students of Ed Jaynes's writings on
probability theory in physics may recognize the last part of my title,
borrowed from a commentary by Jaynes on the history of Bayesian and
maximum entropy ideas in the physical sciences (Jaynes 1993).  This
bit of plagiarism is intended as a homage to Jaynes's influence on this
area---and on my own research and thinking.
\end{acknowledgement}
%



\newcommand{\apj}{Astrophysical Journal\ }
\newcommand{\prd}{Physical Review D\ }
\newcommand{\mnras}{Mon. Not. Roy. Astron. Soc.\ }
\newcommand{\nat}{Nature\ }
\newcommand{\araa}{Annual Review of Astronomy \& Astrophysics\ }



\end{document}